\documentclass[twocolumn,twocolappendix,trackchanges]{aastex6}

\begin{document}

\title{Presence of turbulent and ordered local structure within ICME shock-sheath and its contribution in Forbush decrease}


\author{Zubair Shaikh \altaffilmark{1,2},Anil Raghav\altaffilmark{1*},  Ankush Bhaskar\altaffilmark{2}}

\affil{\altaffilmark{1}University Department of Physics, University of Mumbai, Vidyanagari, Santacruz (E),
Mumbai-400098, India }
\affil{\altaffilmark{2}Indian Institute of Geomagnetism, New Panvel, Navi Mumbai-410218, India.}


\altaffiltext{*}{principle author}
\altaffiltext{@}{raghavanil1984@gmail.com}

\begin{abstract}
The transient interplanetary disturbances evoke short time cosmic ray flux decrease which is known as Forbush decrease. The traditional model and understanding of Forbush decrease suggest that the sub-structure of interplanetary counterpart of coronal mass ejection (ICME) independently contributes in cosmic ray flux decrease.  These substructures, shock-sheath and magnetic cloud (MC) manifest as classical two-step Forbush decrease. The recent work by Raghav et al (2016a) has shown multi-step decreases and recoveries within shock-sheath. However, this can not be explained by ideal shock-sheath barrier model. Further, they suggested that the local structures within the ICME's sub-structure (MC and shock sheath) could explain this deviation of FD profile from the classical FD. Therefore, present study attempts to investigate the cause of multi-step cosmic ray flux decrease and their respective recovery within shock-sheath in detail. 3D-hodogram method has been utilized to get more details of the local structures within the shock-sheath. A 3D-hodogram method unambiguously suggests the formation of small scale local structures within the ICME (shock-sheath and even in MC). Moreover, the method could differentiate the turbulent and ordered Interplanetary Magnetic Field (IMF) regions within the substructures of ICME. The study explicitly suggests that the turbulent and ordered IMF regions within shock-sheath  do influence cosmic-ray variations uniquely.

\end{abstract}


\section{Introduction} \label{sec:intro}
The interplanetary space is filled with cosmic rays. The modulation of cosmic ray eventuate from its interaction with interplanetary magnetic field (IMF). The modulation becomes very prominent when the emission of very large coherent magnetic structure known as interplanetary coronal mass ejection (ICME) take place from the surface of the Sun. The excess speed of ICME over ambient solar wind speed provokes shock-sheath ahead of it. The propagation of this huge magnetic field structure (ICME magnetic cloud (MC)) with shock-sheath in heliosphere induce decrease in cosmic ray flux. This can be demonstrated through the ground based neutron monitors (NMs) around the globe which continuously observe cosmic ray flux \citep{Cane2000,Lockwood1971}. The sudden transient decrease in cosmic ray flux caused by interplanetary disturbances is known as Forbush decrease \cite{Forbush1937}. The corotating interaction region (CIR) and ICME are main drivers of Forbush decrease. The CIR generates recurrent, symmetric and weak Forbush decrease, whereas ICME induced Forbush decrease is non-recurrent, highly asymmetric and strong in nature. Moreover, the magnetic field barrier and solar wind speed are considered to be the main drivers in ICME-associated Forbush decrease \citep{Belov2001,Dumbovic2012,Bhaskar2016a,Bhaskar2016b}.

The past studies of Forbush decrease show either one-step or two-step cosmic ray decrease profile during ICME  transit \citep{Cane2000,Richardson2010,Richardson2011,Richardson1996,Arunbabu2013}. The two-step traditional model and observational studies of Forbush decrease profile manifest that the first step and the second steps are caused by different ICME sub-structures i.e. shock sheath and MC respectively (for example \cite{Anil2014}). However, note that the one-step or the two-step FD profiles depends not only on the structure of the ICME but also on the geometric location of the observer~\citep{Richardson2010}.  This can be explained as, one-step Forbush decrease profile is observed, if observer passes only through the shock-sheath or MC (with weak shock-sheath) and two-step decrease, if the observer passes through both the regions. 

Wibberenz et al. (1998) has put forward the shock-barrier model to understand contribution of the shock-sheath in Forbush decrease.  This model considers the diffusion of cosmic rays across the propagating diffusive barrier. Therefore depression of cosmic ray flux is observed during shock-sheath transit across the Earth \citep{Wibberenz1998}. The one-step/gradual decrease in cosmic ray flux is expected during complete transit of shock-sheath region according to this model. However, observations demonstrate that one-step (corresponding to small part of shock-sheath) or multi-step decreases with respective recovery during its transit, which cannot be explain by this model.

\begin{figure}[!ht]
\includegraphics[width=0.5\textwidth]{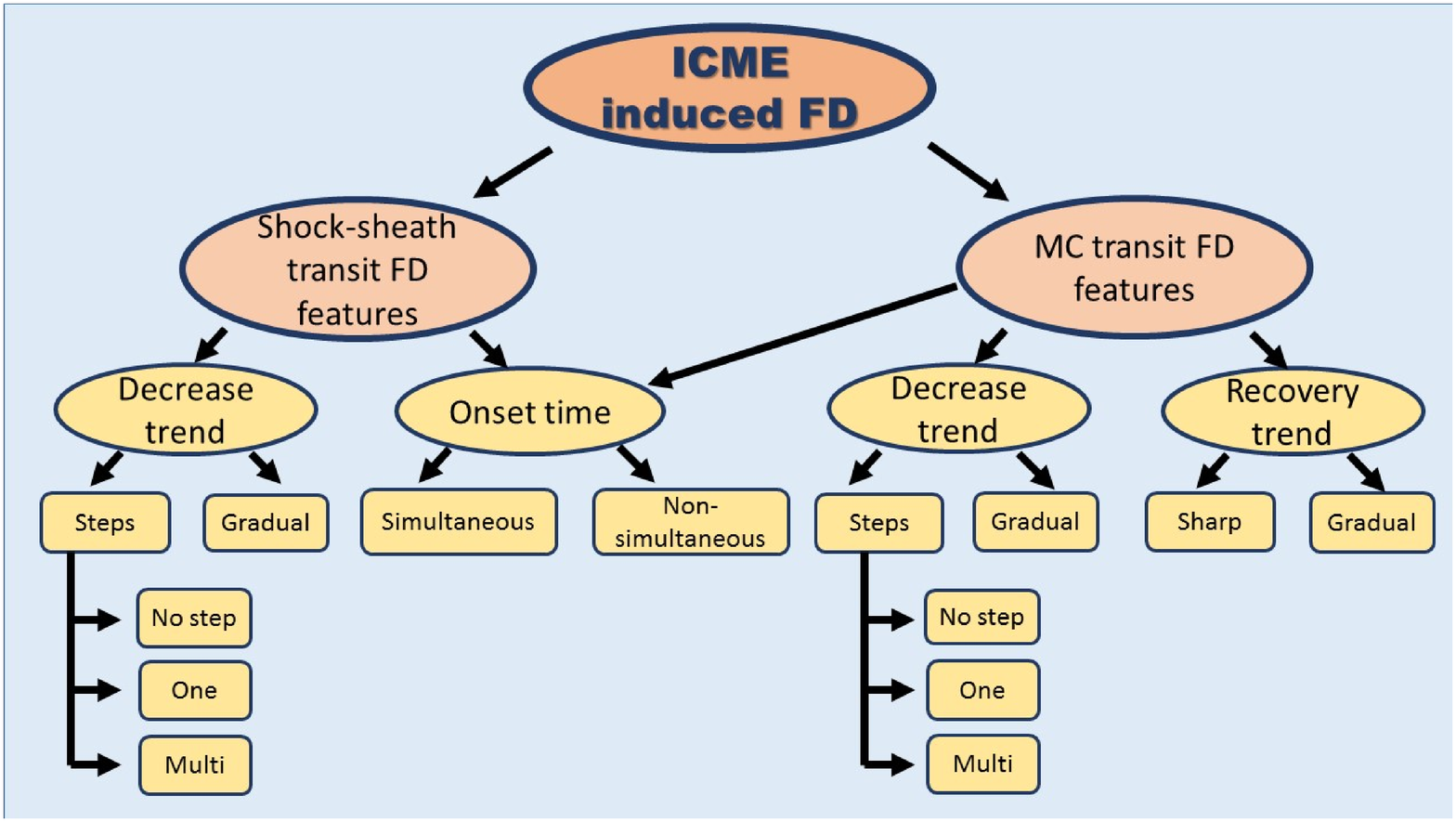}
 \caption{\textit{General features of Forbush decrease profile during ICME transit given in \cite{Raghav2016a}.}}
 \label{fig:7}
 \end{figure}

Interestingly, Jordan et al.(2011) has studied 233 ICME events which could have produced two-step Forbush decrease, but only 13 two-step Forbush decrease events were observed. Therefore, they proposed to discard the classical two-step Forbush decrease model and suggested to consider the possible contribution of local small-scale magnetic structure within ICME in Forbush decrease \citep{Jordan2011}. Moreover, Raghav et al (2016a), has summarized general features of Forbush decrease profile (see in Figure~\ref{fig:7}) and proposed a new classification scheme for Forbush decrease phenomena. Further, they concluded that not only the sub-structures (shock-sheath and MC), but also localized structures within the sub-structure have effective role in Forbush decrease profile \citep{Raghav2016a}. 

In the light of the above discussion, it is important to study cosmic ray variations during shock-sheath and MC transit independently and further evaluate the influence of the local structures within substructures in detail. Thus, the main objective of the present study is to unravel the cause behind the multi-step Forbush decrease and the recovery of cosmic ray flux within the shock-sheath region.

\section{Data and methodology}
To understand the cosmic ray's response to the shock-sheath region, we have analyzed two ICME induced Forbush decrease events (amplitude $\geq 8 \%$) occurred on September 17, 2000 and September 24, 1998 from catalog of Neutron Monitor Database (NMDB at \url{www.nmdb.eu}). The neutron flux data of 51 neutron monitor (NM) observatories are available at \url{nest2.nmdb.eu}. However, of certain events, the data from few laboratories are missing. We have used neutron flux data (with 5 minute time  resolution) retrievable at the above website for studying events. Each neutron monitor observatory has their local characteristics and baseline value of neutron flux. Therefore, we normalized the neutron flux intensity of each observatory. The normalized percentage variation ($\%$) of each Neutron monitor observatory is defined as
\begin{equation}
  {{N_{norm}(t)}={\frac{N(t)-N_{mean}}{N_{mean}}}\times{100}}                           
\end{equation}

where, $N_ {mean}$  is averages of quiet day/days neutron flux of a specific observatory and  $N(t)$ is neutron flux at time $ t $ of the same specific observatory.
We have classified neutron monitor data into three broad energy windows, (i) low rigidity ( 0-2 GV), (ii) Medium rigidity ( 2-4.5 GV ) and (iii) high rigidity ( $\ge$ 4.5 GV). The presented data for each energy band  is the average normalized neutron flux of all observatory comes under given energy band.

To investigate the structure of the ICME (shock-sheath and magnetic cloud topology) of studied events, we have  used one-minute and five-minute time resolution interplanetary  data from OMNI database \url{cdaweb.gsfc.nasa.gov}. The Interplanetary parameters include strength of IMF ($B_{total}$) along with their components $B_x$ , $B_y$ , $B_z$  in (nT), solar wind speed  (km/s), plasma Temperature (K), proton density ($n/cm^3$) and plasma beta.

The boundaries of the shock-sheath and corresponding ICME magnetic cloud are obtained from the Richardson et al. (2010) \citep{Richardson2010}. The Forbush decrease onset is determined using visual inspection, where neutron flux starts to decrease sharply. 

A 2D-hodogram method is used to visualize rotation/structures of IMF in space as well as  in magnetospheric physics. The observation of semi/arc circular pattern in one of the planes, $B_x-B_y$ or $B_x-B_z$ or $B_z-B_y$ is a good indicator of rotational structures present in interplanetary space \citep{Khabarova2015,Khabarova2016}. However, temporal evolution or/and third IMF component contribution in magnetic structure cannot be studied using this 2D-hodogram method. Therefore, we have used 3D-hodogram method to effectively visualize the IMF configuration of different regions within the shock-sheath \cite{Raghav2016b}. To make hodogram a one-second time resolution IMF data from ACE database  \url{www.srl.caltech.edu/ACE/ASC/level2/} is utilized.  


\section{Observations and interpretations}
We have investigated two ICME shock-sheath regions to study the cause of cosmic ray flux decrease and their recovery within shock-sheath. The shock-sheath region transit corresponding to the ICME event occurred on September 17, 2000 (see Figure \ref{fig:1}) exhibiting single step decrease following recovery in cosmic ray flux. Whereas, September 24, 1998 (see Figure \ref{fig:3}) ICME shock-sheath transit demonstrate multi-step decreases \& recoveries. For detail study, we have divided shock-sheath region into sub-parts depending on the cosmic ray flux variations. These sub-parts correspond to decreases and their respective recoveries in cosmic ray flux within the shock-sheath. We have used 3D-hodogram analysis technique to investigate magnetic configuration of each sub-part of ICME shock-sheath.

\subsection{September 17, 2000} 

\begin{figure}[!ht]
\includegraphics[width=0.5\textwidth]{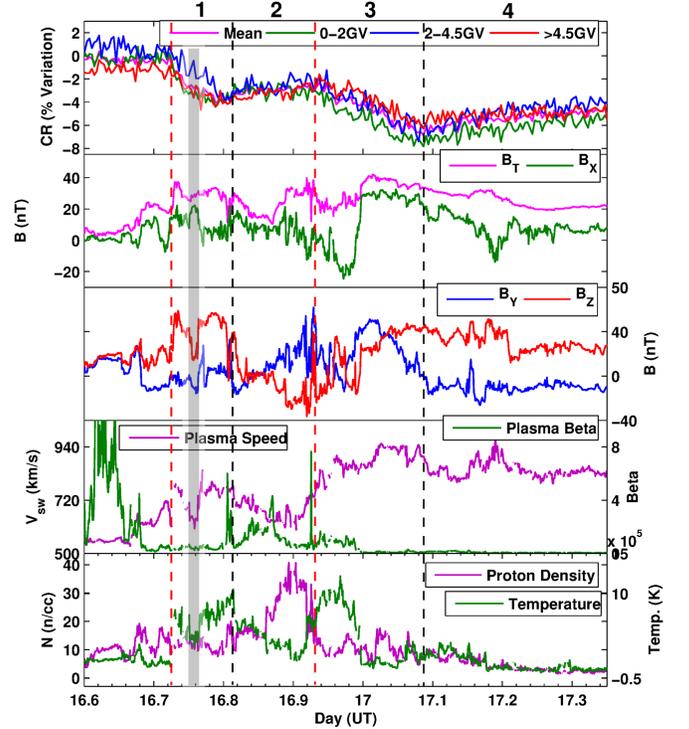}
 \caption{\textit{Forbush decrease event occurred on September 17, 2000. It has five panels, top most panel shows temporal variation of normalized neutron flux with their respective band of rigidities. The $2^{nd}$ and $3^{rd}$ panels show interplanetary magnetic field ($B_{Total}$ \& its $B_X$-component) and ($B_Y$ \& $B_Z$-component) respectively. The $4^{th}$ panel shows solar wind speed \& plasma beta data respectively. The bottom panel shows proton density and plasma temperature variation. The shock-sheath boundaries are shown with red  vertical dash lines. The four different regions of the shock-sheath are separated by  vertical dash lines.}}
 \label{fig:1}
 \end{figure}

The classical two step, simultaneous Forbush decrease event occurred on September 17, 2000 is shown in Figure \ref{fig:1}.  The sudden sharp enhancement in IMF, solar wind speed and proton density  indicate arrival of ICME's shock at the Earth's bow-shock (shown by the first vertical red dash line). The cosmic ray flux shows simultaneous gradual decrease accompanying with the onset of shock. The complete shock-sheath transit takes about $ \sim 5 ~hours$ (region between two vertical red dash line). However, the first step decrease is observed only about $ \sim 2.4 ~hours$. During rest of the shock-sheath crossing, cosmic ray flux shows slow gradual recovery. These observations manifest that only front edge of the shock-sheath (mostly shock-front) contributes in cosmic ray decrease whereas remaining shock-sheath leads to recovery. 

Similarly, after the onset of MC, cosmic ray shows gradual decrease for $\sim 3.8 ~hours$. Moreover, during rest of the MC crossing, cosmic ray flux recovery is observed. These observations suggest that the enhance interplanetary magnetic field strength contributes in decrease during the onset of MC. Whereas gradual decrease in IMF strength give rise to the recovery of cosmic ray flux.
 
 As note earlier, based on cosmic ray variations, we have divided observed region in four local-regions (sub-parts) for this event. First and second local regions correspond to the  decrease and recovery of cosmic ray flux within shock sheath. Whereas, third and fourth local regions correspond to the  decrease and recovery of cosmic ray flux within MC. The boundaries of local regions are indicated  by vertical dash lines which are shown in figure \ref{fig:1}. Further 3D-hodogram have been constructed for better visualization of the IMF configuration during each local region crossing.  
 
 \begin{figure}[!ht]
 \includegraphics[width=0.5\textwidth]{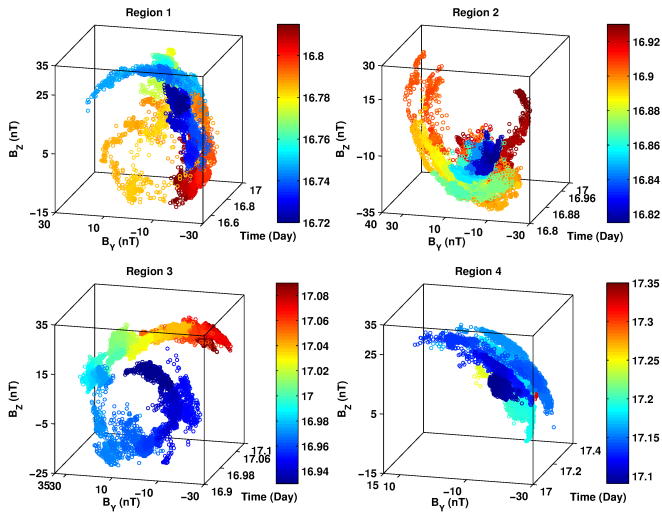}
  \caption{\textit{Four hodograms of the four shock-sheath region as shown in Fig.5. Each hodogram represents the 3D projection of magnetic field in $B_z - B_y $ plane, where the component $B_x$ represents in-out direction. Th colorbars shows the temporal variation of the four shock-sheath regions.
    the top left and right hodogram are for region 1 and 2, while the bottom left and right is for the region 3 and 4 as shown in the Figure 2, respectively.}}
  \label{fig:2}
  \end{figure}
  
  \subsubsection{Region 1}
During crossing of local region 1, the total IMF shows enhancement in total field strength. The all IMF components shows fluctuations though-out the local region 1 except for a small interval of time in which $B_y$ and $B_z$ shows some rotation (see gray shaded strip in Figure~\ref{fig:1}). This could be ascribed as small magnetic island (\cite{Khabarova2015,Khabarova2016,Raghav2016b}) formation within local region 1. The effective visualizations of these observations could be seen in Figure \ref{fig:2}. The left top  hodogram clearly demonstrate that the initial (blue shade) and end part (red shade) of the region 1 is highly fluctuating. Moreover, only small time interval (shown as sky-blue arc) is the evidence of magnetic island formation. In-summary we conclude that the high fluctuation in IMF is due to heating of shock-sheath plasma i.e. turbulence present in local region 1 could be the cause of cosmic ray decrease. 

\subsubsection{Region 2}
The total IMF shows gradual decrease followed by recovery with small fluctuations in local region 2 shown in Figure \ref{fig:1}. Further, the IMF components $B_x$, $B_y$, and $B_z$ show gradual variation with some fluctuations. Beside this, the plasma temperature shows decrease during region 2 transit. Moreover, the right top hodogram of Figure \ref{fig:2} clearly demonstrate the presence of various arc planes. This distinctly indicates the existence of magnetic island within the local region 2. These observations evince that the magnetic structure could be causal to cosmic ray flux recovery.

\subsubsection{Region 3}
In region 3, total IMF and its all components show high fluctuations whereas, plasma temperature and solar wind speed depicts enhancement for about $\sim 1.7 ~hours$. Further sudden increase/decrease in total IMF strength/plasma temperature is observed followed with steady and slow decrease. The left bottom hodogram also demonstrates high fluctuations followed with clear semi-circle structure. These observations imply that the front edge of MC is highly fluctuating i.e. turbulent. However, rest part of region 3 indicate the unambiguous evidence of rotating magnetic structure i.e. ICME flux-rope. These observations reveal that the turbulent region and/or enhanced magnetic field strength region could be responsible for cosmic ray flux decrease.  

\subsubsection{Region 4}
The local region 4 is the part of MC, in which depicts gradual decrease in total IMF with small fluctuations. The solar wind speed, plasma density and temperature shows steady variations. The 3D-hodogram demonstrate the various arc planes which could be ascribed to rotational magnetic structure i.e. the feature of MC.
The cosmic ray flux shows gradual and steady  recovery which indicates that the magnetic ordered structure contributes in cosmic ray recovery.

\subsection{September 24, 1998}

\begin{figure}[!ht]
\includegraphics[width=0.5\textwidth]{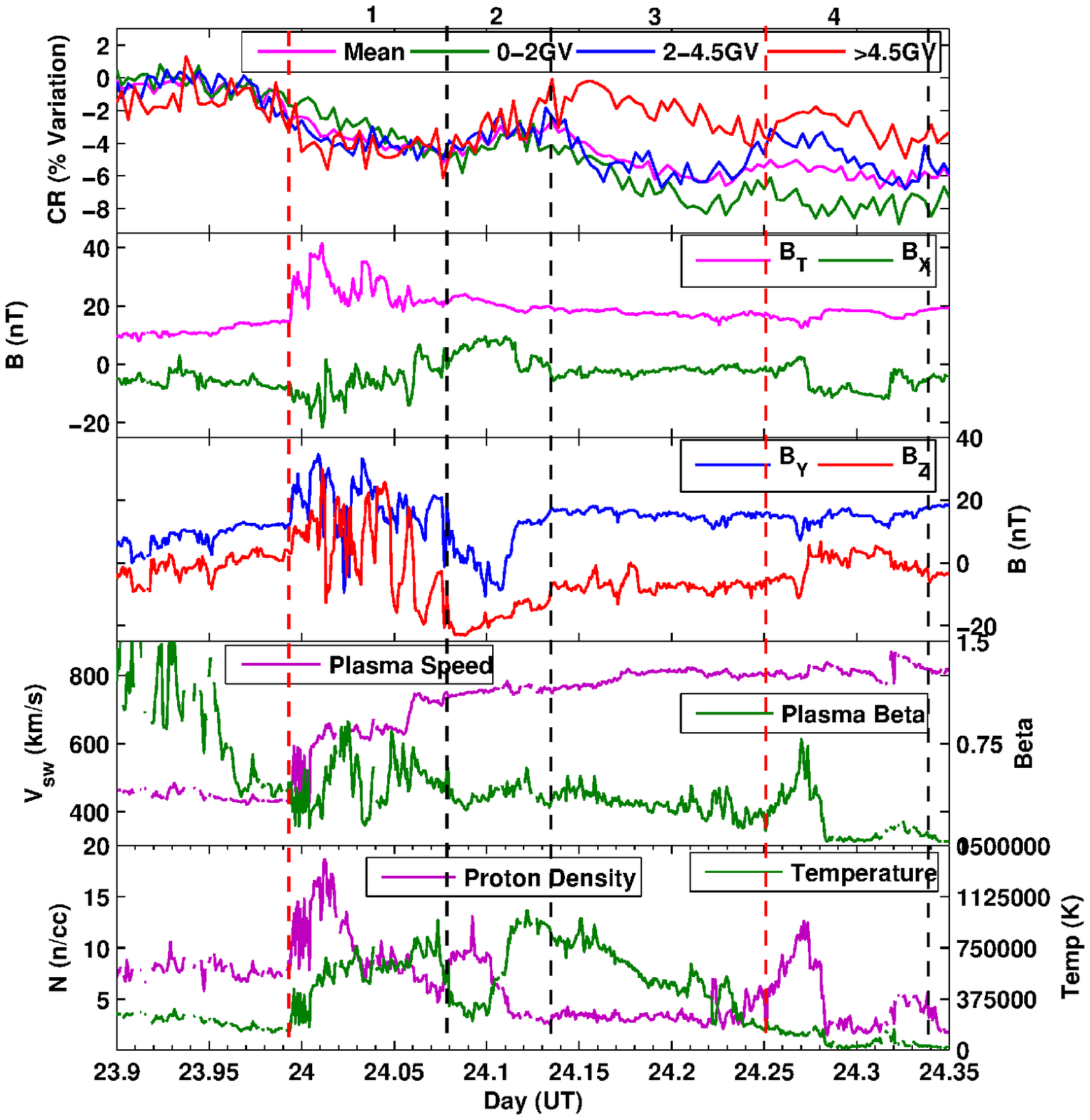}
 \caption{\textit{Forbush decrease event occurred on Septenber 24, 1998. It has five panels, top most panel shows temporal variation of normalized neutron flux with their respective band of rigidities. The $2^{nd}$ and $3^{rd}$ panels show interplanetary magnetic field ($B_{Total}$ \& its $B_X$-component) and ($B_Y$ \& $B_Z$-component) respectively. The $4^{th}$ panel shows solar wind speed \& plasma beta data respectively. The bottom panel shows proton density and plasma temperature variation. The shock-sheath boundaries are shown with red  vertical dash lines. The four different region of the shock-sheath are separated by vertical dash lines.}}
 \label{fig:3}
 \end{figure}

\begin{figure}[!ht]
 \includegraphics[width=0.5\textwidth]{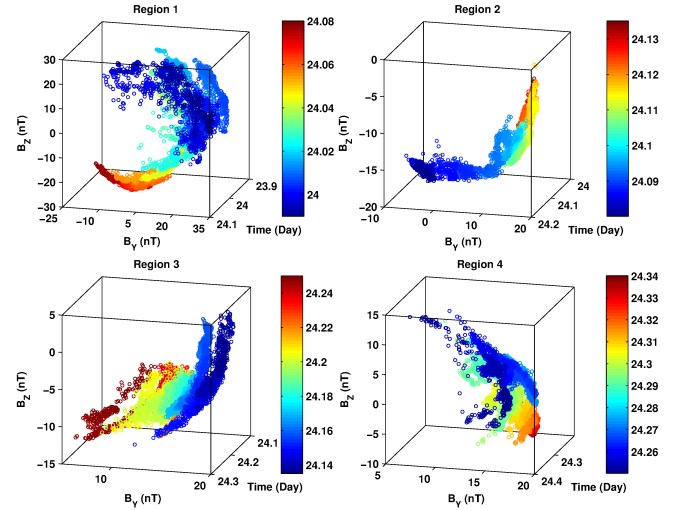}
  \caption{\textit{ Four hodograms of the four shock-sheath region as shown in Fig.4. Each hodogram represents the 3D projection of magnetic field in $B_z - B_y $ plane, where the component $B_x$ represents in-out direction. Th colorbars shows the temporal variation of the four shock-sheath regions. The top left and right hodogram are for region 1 and 2, while the bottom left and right is for the region 3 and 4 as shown in the Figure 4, respectively.}}
  \label{fig:4}
  \end{figure}

The complex multi-step Forbush decrease event occurred on September 24, 1998, is shown in Figure \ref{fig:3}. It shows two-step decrease in cosmic ray flux with recovery within the shock-sheath region of ICME. The commencement of shock at the $Earth’s$ bow-shock (shown by first vertical red dash line) can be identified by sudden sharp enhancement in IMF, solar wind speed and proton density. The complete shock-sheath transit takes about ∼$6.5 ~hours$ (region between two vertical red dash line). The cosmic ray flux starts decreasing well before the commencement of the shock. This might be due to the  enhanced magnetic field before the commencement of the shock. The effective visualization of IMF structure  within the shock-sheath region is demonstrated using 3D-hodogram method. Figure~\ref{fig:4} shows 3D-hodograms for all the sub-regions of Figure~\ref{fig:3}.

\subsubsection{Region 1}
 The total IMF, solar wind speed, plasma temperature, and plasma density is significantly enhanced which is signature of compressed and heating nature in region 1. In region 1, a strong stochastic fluctuations are observed in total IMF and its components. This can also be clearly seen in top left hodogram in Figure\ref{fig:4}. The hodogram illustrates that the front part of the region 1 is highly fluctuating (see blue shed). These observations  suggest that the front edge of shock sheath (shock-front) is highly turbulent and significantly contributing in cosmic ray flux decrease. 

\subsubsection{Region 2}
During region 2 transit, the total IMF shows steady slow decrease, while its components have smooth variations. A clear  orientated structure (rotational) is observed, especially in $B_y $ and $B_Z$ components of the IMF. The proton density and plasma temperature increase/decrease in this region, while the solar wind speed shows slow  gradual increase. All these observations can be interpreted as the formation of rotational structure within the shock-sheath region. The 3D-hodogram of the local region 2, is represented at the top right in the Figure \ref{fig:4}. We can clearly observe the rotational structure (semicircle) of the IMF in y-z plane. This rotational structure/flux-rope within the local region 2 of the ICME shock-sheath seems responsible for the observed recovery of cosmic ray flux.

\subsubsection{Region 3}
During second step decrease (i.e. local region 3), a steady variation in the total IMF and its components $B_X$ and $B_Y$ is observed, while the $B_Z$ shows small fluctuation. The plasma density, solar wind speed, plasma beta shows steady variation, whereas the plasma temperature decreases gradually with small fluctuations. The circle arc planes with different arc-length are observed in the bottom left 3D-hodogram. Note that the end part of region 2 has different arc length as compared to that of initial part of region 3. However, the orientation of these arc planes is similar. This indicate that the rotational structures in local region 2 may be extended in local region 3.  Interestingly, in local region 3 IMF is steady (non-decreasing strength) and has down to dusk orientation. This is clear evidence of flux-rope formation in shock-sheath region. The steady and dawn to dusk oriented IMF in flux-rope (observed in region 3) 
could be responsible for the slow and gradual decrease in cosmic ray due to diffusion process.

 
\subsubsection{Region 4}
Further, in local region 4, total IMF shows small enhancement with steady variations, however, all its components depicts small fluctuations superposed over the steady variations. The solar wind speed and plasma temperature shows steady variations while the proton density and plasma beta shows gradual enhancement followed with sudden decreases to its ambient value. The right bottom hodogram also depicts that the initial part of region 4 is fluctuating but rest of it shows oriented arc plane with different length. This region also has similar rotational structure (flux-rope) as observed in region 2 and 3 respectively except at the initial part i.e. at the transition region between shock-sheath and MC. Moreover, during this transit, we have observed overall gradual decrease in cosmic ray flux. The enhanced IMF field strength, along with the fluctuating IMF components could result in observed cosmic ray flux decrease.

\section{Discussion \& conclusion}
The Forbush decrease phenomenon is interpreted  on the basis of diffusion of cosmic rays in convecting interplanetary magnetic field structures. These structures generally consist of two broad sub-structures, shock-sheath and flux rope (MC).  The shock-sheath is mainly characterized by presence of stochastic fluctuations of magnetic field, whereas MC is identified as an ordered magnetic structure  \citep{Cane2000, Richardson2011, Anil2014}.
In general, the first step of forbush decrease is generally ascribed to shock-sheath and second one is to MC \citep{Wibberenz1998,Cane2000,Arunbabu2013,Anil2014}. To be precise, the model given by \cite{Wibberenz1998},  which is based on diffusion of cosmic rays across the propagating diffusive barrier, is widely used to explain cosmic ray decrease during shock-sheath crossing. The one-step/gradual decrease in cosmic ray flux is the direct implication of this model. However, this model does not clarify the origin of multi-step decrease and/or recovery during shock-sheath transit. The one-step, the two-step and the multi-step complex events have been reported and suggests led to consider the contribution of local structures within the sub-structures in Forbush decrease profile \citep{Jordan2011, Raghav2016a}. But, Which part of the shock-sheath or MC contributes to step/gradual decrease or/and in recovery of cosmic ray flux has remained as an open problem. Here, we have carried out detailed investigation of the local magnetic structures within shock-sheath region and their contribution in cosmic ray decrease and/or recovery.

\begin{figure}[!ht]
 \includegraphics[width=0.5\textwidth]{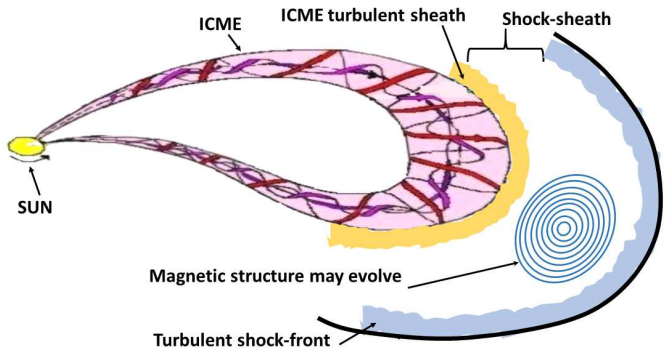}
  \caption{\textit{Schematic diagram of ICME evolution in interplanetary space.}}
  \label{fig:6}
  \end{figure}

To address this issue, we have examined two large Forbush decrease events occurred on September 17, 2000 and September 24, 1998. The first event shows only one-step decrease with recovery and second event shows two-step decrease with respective recoveries within the shock-sheath region. The 3D-hodogram method used in this work gave effective visualizations of magnetic configuration of the local structures formed within the shock-sheath region. Moreover, this method explicitly differentiate the turbulent and ordered IMF configuration. 

This study clearly demonstrates that  (i) the random fluctuations (i.e turbulence) and strength of IMF contributes in the cosmic ray decrease,  (ii) the ordered structure with almost constant magnetic field participates in cosmic ray flux decrease, (iii) the ordered structure with declining magnetic field strength contributes in cosmic ray flux recovery.  The generally accepted view is that the shock-front is turbulent due to plasma compression and heating which results in cosmic ray flux decrease. The presence of the  ordered structure ( magnetic island/flux-rope) in the shock-sheath is intriguing and their origin is unclear at present. There exists another possibility, that the ordered structure (magnetic island/flux-rope) evolves in the later part of the shock-sheath region as a consequence of plasma relaxation process~\cite{Tylor1986}. The shock-sheath plasma is not perfectly conductive due to the presence of turbulence and hence, can reach to Taylor state (low potential energy) by forming magnetic island like structures within shock-sheath. This newly evolved ordered structure within shock-sheath can contribute to recovery of cosmic ray flux. Further, as the MC reconnection with shock-sheath may give rise to turbulent conditions which is in transition region, occurring between the sheath and MC. This process can be thought as follows: (i) Magnetic reconnection give rise to electric field. (ii) Electric field accelerates charged particles. (iii) Acceleration process increases the kinetic energy. Moreover, it should also increase the plasma temperature (Signature of this is evident in Figure \ref{fig:1}). (iv) The net effect of this is the formation of turbulent region just before the MC. The possible visualization using artistic picture of ICME (shock-sheath and MC) and evolved magnetic structure within shock-sheath is shown in Figure~\ref{fig:6}.  In summary, observations illustrate that the transition regions, which are in general turbulent, and the front edge of MC, with enhanced IMF strength contributes in cosmic ray decrease. Although, rest of the MC contributes in recovery of cosmic ray flux, due to the gradual decrease in magnetic field strength. It is also possible that as ICME travels in interplanetary space, it drags away the already present magnetic island in solar wind. The shock-sheath material of ICME is then contaminated by these structures and, therefore, magnetic island might appear as a part of the ICME. These kind of structures may also originate from ICME-ICME, ICME-CIR, and/or ICME-solar wind interactions. 
  
 To conclude, the study explicitly shows that turbulent as well as ordered IMF regions do exist in shock-sheath region. These ordered and turbulent regions affect cosmic ray flux variations distinctively.

\section{Acknowledgement}
We acknowledge the NMDB database (www.nmdb.eu) founded under the European Union's FP7 programme (contract no. 213007). We are also thankful to all neutron monitor observatories listed on website. We are thankful to CDAWeb and ACE science center for making interplanetary data available. We are thankful to Department of Physics (Autonomous), University of Mumbai, for providing us facilities for fulfillment of this work. Authors would also like to thank Aditi Upadhyay for their valuable English correction.\\



\end{document}